\documentclass[graphicx,aps,pre,epsfig,showpacs]{revtex4}
\usepackage{graphicx}
\usepackage{bm}
\usepackage{amsmath}
\begin{document}
\title{Metastable configurations on the Bethe lattice}

\newcommand{\be}{\begin{equation}}
\newcommand{\bea}{\begin{eqnarray}}
\newcommand{\ee}{\end{equation}}
\newcommand{\eea}{\end{eqnarray}}
\newcommand {\lan} {\langle}
\newcommand {\ran} {\rangle}
\def\(({\left(}
\def\)){\right)}
\def\[[{\left[}
\def\]]{\right]}
\def\bi{\bibitem}
\def \form#1 {eq. (\ref{#1}) }
\def \parziale#1#2  {{\partial {#1} \over \partial {#2}}}
\def \atanh{\hbox{atanh}}
\def \cM{{\cal M}}
\def \cN{{\cal N}}
\def \cC{{\cal C}}
\def \cH{{\cal H}}
\def \cP{{\cal P}}
\def \cQ{{\cal Q}}
\def \cR{{\cal R}}
\def \cV{{\cal V}}
\def \cL{{\cal L}}
\def\la{\langle}
\def\ra{\rangle}
\def \Tr {\mbox{Tr}}
\def \ba#1 {\overline{#1}}
\def  \bh {{\bf h}}
\def  \bg {{\bf g}}
\def  \bF {{\bf F}}
\def  \si {\sigma}
\def  \s {\sigma}
\def \vp {{\vec p}}
\author{A. Pagnani}
\affiliation{Dipartimento di Fisica, SMC, INFM, Universit\`a di Roma 1 
{\em La Sapienza, P.le A. Moro, 2 -- 00185 Roma, Italy.}}
\author{G. Parisi}
\affiliation{Dipartimento di Fisica, SMC, INFM, and INFN, Universit\`a
di Roma 1 {\em La Sapienza, P.le A. Moro, 2 -- 00185 Roma, Italy.}} 
\author{M. Rati\'eville}
\affiliation{Dipartimento di Fisica, SMC, INFM, Universit\`a di Roma 1
{\em La Sapienza, P.le A. Moro, 2 -- 00185 Roma, Italy}} 
\affiliation{Laboratoire de Physique Th\'eorique et Mod\`eles Statistiques
Universit\'e Paris Sud -- {\em 91405 Orsay, France.}}

\begin{abstract}
We present a general analytic method to compute the number of metastable
configurations as a function of the energy for a system of interacting
Ising spins on the Bethe lattice. Our approach is based on the cavity
method. We apply it to the case of ferromagnetic interactions, and also to
the binary and Gaussian spin glasses. Most of our results are
obtained within the replica symmetric ansatz, but we illustrate how
replica symmetry breaking can be performed.
\end{abstract}

\maketitle

\section{Introduction}

Despite years of efforts, the nature of the glassy phase of finite
dimensional spin glasses is still not clear. It is still debated
whether the replica symmetry breaking (RSB) scheme proposed by Parisi
to solve the mean field fully connected Sherrington-Kirkpatrick model
(SK), and implying the existence of an exponentially large number of
pure states with an ultrametric structure, holds in some way for these
systems. Part of the difficulty to settle the question is the very
poor analytical tractability of finite dimensional systems.

Looking for more realistic -- but still tractable -- models than SK,
much attention was recently paid to spin glasses on random graphs of
finite connectivity. These models incorporate the short range nature
of interactions, but without the underlying geometry of the finite
dimensional models. It has been shown that the replica symmetric (RS)
solution of spin glasses on random graphs of finite connectivity is
unstable at low temperature \cite{mottishaw,sompo}. Unfortunately working out the RSB scheme
is far more difficult than for the SK, because the glassy phase can no
longer be characterized by the only two spin overlap $\la \s_a \s_b
\ra$ between two distinct replicas $a$ and $b$, but requires all of
the multi spin overlaps.  Recently a new approach was suggested in
\cite{mezpar}, based on the cavity method and population algorithms,
that allows for a numerical solution at the level of one step RSB
(1RSB). The method can be virtually extended to any step of RSB, at
the price of increasing computer resources.

In this paper, we use similar ideas to shed a new light onto an old
problem: the computation of the number of metastable configurations. A
configuration is said to be metastable if its energy cannot be
decreased by flipping a single spin. One expects, within the RSB
scenario, that a consequence at zero temperature of the exponential
number of pure states, is an exponential number of metastable
configurations. Beware however that these two concepts are not as
obviously linked as it might seem: in systems with finite connectivity
the zero temperature limit of the pure states are not the metastable
configurations, but are argued to be the configurations stable with respect to an
arbitrary finite number of spin flips \cite{birolimonasson,PARISI_MEZARD_T0}.

Let us call configurational entropy $S_C(E)$ the logarithm, divided by
the number of spins, of the number of metastable configurations having
an energy density equal to $E$. Lots of efforts have been devoted to
the study of $S_C$ for SK. More recently, attention turned to spin
systems on random graphs. On random graphs with fixed finite
connectivity, annealed computations were carried out for the binary
spin glass \cite{dean}, and the ferromagnet was addressed in
\cite{lefevredean}. On random graphs of fluctuating finite
connectivity, the case of the ferromagnet was solved in \cite{berg} --
also {\sl via} a population algorithm, but in a different context as
ours --, and the authors give hints how to address the quenched
computation of spin glasses.

Our method, which is quite general, enables us to recover all the
above results. As new contributions, we carry out quenched
computations in the case of the Gaussian or binary spin glass on
random graphs of fixed finite connectivity, and we exemplify the
implementation of 1RSB.

In the following we stick to the case of random graphs of fixed finite
connectivity. For the sake of concision, we display no result about
the fluctuating connectivity, but it is easy to generalize our method
to this case.

The layout of the paper is the following. In section \ref{definitions}
we give several definitions and we set the notations. In section
\ref{cavity} we derive our basic equations in the RS framework.  In
section \ref{sec_fm} we analytically solve these equations in the case
of a ferromagnet. We compare our results with the microcanonical
approach of \cite{lefevredean}.  In section \ref{binary} we turn to
the binary spin glass, where the coupling constants are $\pm 1$. The
population algorithm shows up there.  In section \ref{gaussian} we
address the slightly more involved Gaussian spin glass.  Eventually,
in section \ref{1RSB} we show how to perform 1RSB, and illustrate the
algorithm on the case of the spin glass with binary couplings.

\section{The system under study}
\label{definitions}

\begin{figure}
\begin{center}
\includegraphics[angle=0,width=14cm]{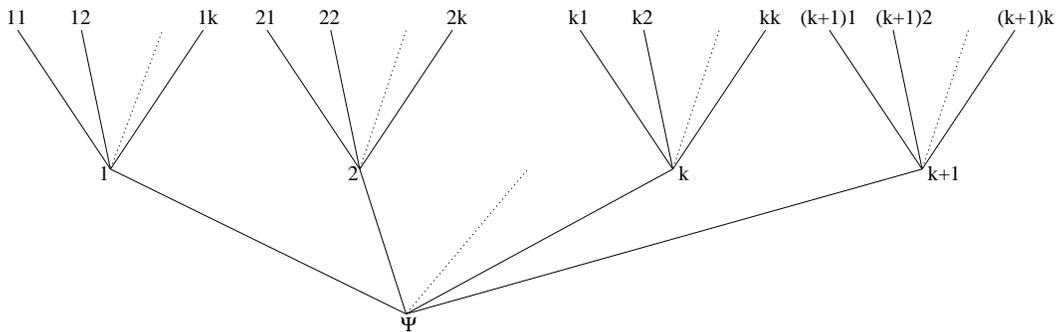}
\caption{Part of a Cayley tree}
\label{mergingkp1}
\end{center}
\end{figure}

Following \cite{mezpar}, we call Bethe lattice a random graph with
fixed connectivity equal to $k+1$, i.e. the number of edges incident
to each vertex is exactly $k+1$. Locally such a graph has the
structure of a Cayley tree, and can be obtained with the following
procedure: starting from a root (or ancestor) $\Psi$, one builds a
first generation of $k+1$ sons, and then successive generations of $k$
sons to each vertex, as displayed in Fig.(\ref{mergingkp1}). By
contrast to the Cayley tree, the Bethe lattice has loops, but small
ones are rare: the typical length of a loop is of order $\log N$. In
practical case one uses the tree-like structure to write down
recursive (or 'cavity' in the language of the physics of disordered
systems \cite{mpv}) equations, and the existence of loops is enforced
by saying that all the spins are equivalent, which is not true for the
Cayley tree (in particular the Cayley tree has strong boundary
effects).

On each vertex $A$ of the Bethe lattice stands a spin, and these spins
interact through the Hamiltonian
\begin{equation}
\label{hamiltonian}
\cH=-\sum_{\{ A,B \}} J_{A,B} \s_A \s_B,
\end{equation}
where the sum runs over all the edges. The coupling constants $J_{A,
B}$ may be fixed, like in the case of the ferromagnet, or they may be
quenched random variables in the case of a spin glass.

The local field acting on the spin $A$ is $H_A=\sum_{B\in\cV(A)}
J_{A,B}\s_B$, where the sum runs over the $k+1$ first neighbors of
$A$. The metastable states are characterized by $\forall A, H_A \s_A
\geq 0$. Our aim is to compute the partition function restricted to
the metastable configurations:

\begin{equation}
Z=\Tr \[[ e^{-\beta \cH}\prod_{A} \Theta (H_A \s_A) \]],
\end{equation}
where $\Theta$ is the Heaviside step-function such that $\Theta(u)=1$
if $u \geq 0$, 0 otherwise, and Tr stands for the summation over all
the possible values of all the spins.

\section{The cavity equations}
\label{cavity}

Let us introduce, in close analogy with the construction of the {\em
infinite tree} in \cite{ALDOUS}, a simple labeling rule for the
vertices of a tree: the sons of the root are labeled $1,2,\dots
k$. Then recursively the sons of a vertex labeled $(i_1\dots i_p)$
are labeled $(i_1\dots i_p\, 1),\dots (i_1\dots i_p\, k)$. If
$A=(i_1\dots i_p)$, we can define the {\em norm} of the symbol as
$|A|\equiv p$.

The basic building block of the cavity solution is a branch of the
Cayley tree, i.e. a subtree made of a given vertex different from the
root and all its descendants. On Fig.~(\ref{tree}) is drawn a
branch rooted at $\Phi$. Let us call it $T_\Phi$.

The key property of a branch is that it is structurally similar to
each of its sub-branches made of a given vertex and all its
descendants. This makes a recursive approach possible to study the
thermodynamics of a branch. We will see afterwards how to use the
results for the Bethe lattice.

\subsection{Computing the metastable states on a branch}

As it will be clear later, on a branch it is convenient to require
that all the spins are stables, except the root. The Hamiltonian of a
branch is

\begin{equation}
\label{Hbranch}
\cH_\Phi=-\sum_{\{ A, B \} \in T_\Phi} J_{A,B} \s_A \s_B.  \ee
The definition of the partition function is
\begin{equation}
Z_\Phi=\Tr \[[ e^{-\beta \cH_\Phi}\prod_{A\in T_\Phi\backslash \Phi} \Theta (H_A \s_A) \]].
\end{equation}
By contrast to all the other spins in $T_\Phi$, which have $k+1$
neighbors, spin $\Phi$ lacks one neighbor. In all the following the
local field acting on such a spin with only $k$ neighbors will be
called a 'cavity' field and denoted by a small $h$ (here $h_\Phi$). We
call $p_\Phi (h,\s)$ the joint density probability of $h_\Phi$ and the
value of spin $\Phi$. One has

\begin{equation}
\label{prima}
p_\Phi(h,\s)=\frac{1}{Z_\Phi} \Tr \[[ e^{-\beta \cH_\Phi}\delta (\s_\Phi-\s)\prod_{|A|\geq 1} \Theta (H_A \s_A)\delta(h_\Phi-h) \]].
\end{equation}
We call $T_1,\dots T_k$ the subtrees of $T_\Phi$ engendered by points
$1,\dots, k$. We can define their Hamiltonians similarly to
Eq.~(\ref{Hbranch}), and write
\begin{equation}
\cH_\Phi=\sum_{i=1}^k \cH_i-h_\Phi \s_\Phi.
\end{equation}
Splitting the Tr, equation (\ref{prima}) can be restated as
\begin{equation}
\label{1}
p_\Phi(h,\s)=\frac{1}{Z_\Phi}  e^{\beta h\s}\, \mathop{\Tr}_{\s_1,\ldots \s_k} \, \delta(h_\Phi-h) \, \mathop{\Tr}_{|A|\geq 2}\, e^{-\beta  \sum_{i=1}^k \cH_i} \prod_{i=1}^k \Theta (H_i \s_i) \prod_{|A|\geq 2} \Theta (H_A \s_A).
\ee
Now separate $H_i$ into two contributions: 
\begin{equation}
\label{2}
H_i=\sum_{j=1}^k J_{i,ij} \s_{ij} +J_{\Phi,i} \s_\Phi,
\end{equation}
and enforce the fact that the first one is the cavity field acting on
spin $i$ in the absence of $\Phi$ through the identities
\begin{equation}
\label{3}
1=\int d h_i \, \delta\Big(\sum_j J_{i,ij} \s_{ij}-h_i\Big).
\end{equation}
Plugging (\ref{2}) and (\ref{3}) into (\ref{1}) one gets
\bea
p_\Phi(h, \s)=\frac{1}{Z_\Phi} e^{\beta h\s}\, \mathop{\Tr}_{\s_1,\ldots \s_k} \, \delta(h_\Phi-h) \prod_{i=1}^k \int d h_i \, \Theta \[[(h_i+J_{\Phi,i} \s)\s_i\]]
\\\nonumber \times \mathop{\Tr}_{|A|\geq 2}\,e^{-\beta  \sum_{i=1}^k \cH_i}\prod_{|A|\geq 2} \Theta (H_A \s_A) \prod_{i=1}^k \delta (\sum_j J_{i,ij}\s_{ij}-h_i)
\eea
Comparing to Eq.~(\ref{prima}), it turns out that the second line in the above equation is nothing but the product of $Z_i p_i(h_i,\s_i)$ for $i=1,\dots k$. So one eventually obtains the following recursion relation:
\begin{equation}
\label{iteration}
p_\Phi(h,\s) =\frac{Z_1\dots Z_k}{Z_\Phi} e^{\beta h\s}\, \mathop{\Tr}_{\s_1,\ldots \s_k} \, \delta(h_\Phi-h) \prod_{i=1}^k \int d h_i \, \Theta \[[(h_i+J_{\Phi,i} \s)\s_i\]] p_i(h_i,\s_i).
\end{equation}
By averaging over the coupling constants and the random graphs (operation denoted by $\overline{\cdots}$), one defines the probability distributions $\cP_\Phi^{(h,\s)}(p)=\overline{\delta(p_\Phi(h,\s)-p)}$. Equation (\ref{iteration}) induces a functional relation between $\cP_\Phi^{(h,\s)}$ and the analogous probability distributions $\cP_i^{(h,\s)}$ for $i=1,\dots k$. In the Bethe lattice all spins are required to be equivalent so we impose the condition
\begin{equation}
\label{selfcon}
\forall i, \,\,\,\, \cP_\Phi=\cP_i=\cP.
\end{equation}
This yields a self-consistency equation for $\cP$.

\begin{figure}
\begin{center}
\includegraphics[angle=0,width=10cm]{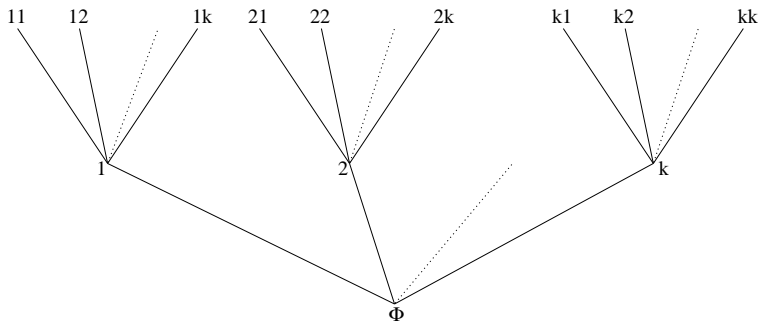}
\caption{A branch}
\label{tree}
\end{center}
\end{figure}

\subsection{Performing measures on the Bethe lattice}

\begin{figure}
\begin{center}
\includegraphics[angle=0,width=10cm]{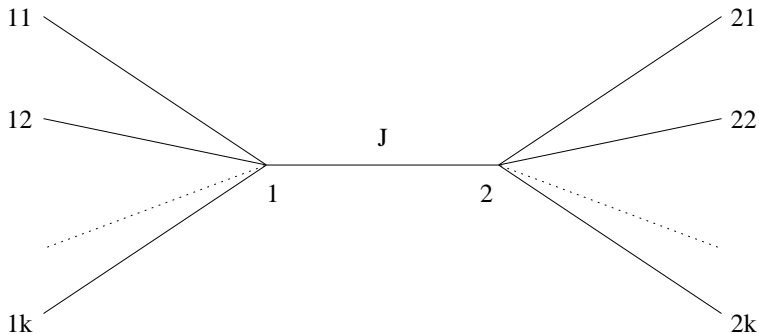}
\caption{Result of the merging (M2) of 2 branches}
\label{merging2}
\end{center}
\end{figure}

Having solved the thermodynamics of a branch, the idea of
\cite{mezpar} is to describe the Bethe lattice as the result of the
merging of several branches. We define two merging procedures:
\begin{itemize}
\item[(M1)] Take $k+1$ branches, with roots labeled $i=1,\dots k$ and
merge them onto a new spin $\Psi$. The resulting tree is the one of
Fig.~\ref{mergingkp1}.
\item[(M2)] Take 2 branches, with roots labeled 1 and 2, and merge
them {\em via} a new bond of coupling constant $J$, without adding any
site. See Fig.~\ref{merging2}.
\end{itemize}

Operation (M1) is very similar to the iteration Eq.~(\ref{iteration}):
the differences are that $k$ is to be substituted by $k+1$, and one
must enforce the stability of the spin $\Psi$. After the merging, the
joint density probability of the value of spin $\Psi$ and the local
field $H_\Psi$ acting on it is $P_\Psi(H,\s)$, given by:
\begin{equation}
P_\Psi(H,\s) =\frac{Z_1\dots Z_{k+1}}{Z} e^{\beta H\s}\Theta(H\s)\, \mathop{\Tr}_{\s_1,\ldots \s_{k+1}} \, \delta(H_\Psi-H) \prod_{i=1}^{k+1} \int d h_i \, \Theta \[[(h_i+J_{\Psi, i} \s)\s_i\]] p_i(h_i,\s_i).
\end{equation}
Summing over all the possible $(H,\s)$ and taking the logarithm one gets the variation of the free energy of the system during the merging:
\begin{eqnarray}
\nonumber \Delta F_1&=&-\frac{1}{\beta} \ln \[[ \sum_{\s}\int dH e^{\beta H\s}\Theta(H \s)\, \mathop{\Tr}_{\s_1,\ldots \s_{k+1}}\, \delta(H_\Psi-H) \prod_{i=1}^{k+1} \int d h_i \, \Theta \[[ (h_i+J_{\Psi, i} \s)\s_i\]] p_i(h_i,\s_i) \]]
\\\label{dF1} &=&-\frac{1}{\beta} \ln \[[\mathop{\Tr}_{\s,\s_1,\ldots \s_{k+1}} \,e^{\beta H_\Psi\s} \Theta(H_\Psi \s) \prod_{i=1}^{k+1} \int d h_i \, \Theta \[[(h_i+J_{\Psi, i} \s)\s_i\]] p_i(h_i,\s_i) \]].
\end{eqnarray}

As far as operation (M2) is concerned, it is straightforward to derive
the variation of the free energy 
\begin{equation}
\label{dF2}
\Delta F_2=-\frac{1}{\beta} \ln \[[\mathop{\Tr}_{\s_1,\s_2} \,e^{\beta J \s_1 \s_2} \int d h_1  d h_2\, \Theta \[[(h_1+J \s_2)\s_1\]]\, \Theta \[[ (h_2+J \s_1)\s_2\]] p_1(h_1,\s_1) p_2(h_2,\s_2)\]].
\end{equation}

One can easily deduce the density of free energy $F$ of the Bethe
lattice from $\Delta F_1$ and $\Delta F_2$. Assuming that each branch
has a number of spins equal to $N$, the system after the merging (M1)
has $N(k+1)+1$ spins, and its free energy can by written in two ways:

\begin{equation}
(N(k+1)+1)F=(k+1) F_{branch}+\overline{\Delta F_1},
\end{equation}
where $F_{branch}$ is the free energy of one of the branches before
the merging. For operation (M2), one has 
\begin{equation} 
(2N)F=2
F_{branch}+\overline{\Delta F_2}.  
\end{equation}

Elimination of $F_{branch}$ between the above equations yields

\begin{equation}
\label{F}
F=\overline{\Delta F_1}-\frac{k+1}{2}\overline{\Delta F_2}.
\end{equation}

Note that the above derivation holds for any extensive and
self-averaging observable. Thus the density of energy $E$ of the Bethe
lattice is

\begin{equation}
\label{E}
E=\overline{\Delta E_1}-\frac{k+1}{2}\overline{\Delta E_2},
\end{equation}
where $\Delta E_1$, resp. $\Delta E_2$, is the variation of energy under the merging process (M1), resp. (M2):
\begin{eqnarray}
\nonumber \Delta E_1 &=&- \sum_\s \int dH P_\Psi(H,\s) H \s
\\\label{dE1} &=& - \frac{\mathop{\Tr}_{\s,\s_1,\dots \s_{k+1}} H_\Psi \s \,e^{\beta H_\Psi\s} \Theta(H_\Psi \s) \prod_{i=1}^{k+1} \int d h_i \, \Theta \[[(h_i+J_{\Psi, i} \s)\s_i\]] p_i(h_i,\s_i)}{\mathop{\Tr}_{\s,\s_1,\dots \s_{k+1}} \,e^{\beta H_\Psi\s} \Theta(H_\Psi \s) \prod_{i=1}^{k+1} \int d h_i \, \Theta \[[(h_i+J_{\Psi, i} \s)\s_i\]] p_i(h_i,\s_i)},
\end{eqnarray}
and
\begin{equation}
\label{dE2}
\Delta E_2=-J \frac{\mathop{\Tr}_{\s_1,\s_2} \,\s_1 \s_2 e^{\beta J \s_1 \s_2} \int d h_1  d h_2\, \Theta \[[(h_1+J \s_2)\s_1\]]\, \Theta \[[ (h_2+J \s_1)\s_2\]] p_1(h_1,\s_1) p_2(h_2,\s_2)}{\mathop{\Tr}_{\s_1,\s_2} \,e^{\beta J \s_1 \s_2} \int d h_1  d h_2\, \Theta \[[(h_1+J \s_2)\s_1\]]\, \Theta \[[ (h_2+J \s_1)\s_2\]] p_1(h_1,\s_1) p_2(h_2,\s_2)}.
\end{equation}
These results can be applied to a wide variety of models: up to now we have not specified the coupling constants.

\section{The ferromagnetic case: An exactly solvable model}
\label{sec_fm}

First, we concentrate on a simple case where all the above equations
can be worked out analytically. The previous formalism allows for the
computation of the metastable states of the $k=2$ Ising
ferromagnet. When the couplings all have the same absolute value
$|J|=1$, the cavity field can take only value in the set
$(-2,0,2)$. Thus integration with respect to the $h_i$ in
Eq.~(\ref{iteration}) and other formulas reduces to a finite
summation. Moreover, in the case of the Ising ferromagnet ({\em
i.e.}~$J=1$) the system is homogeneous, therefore the cavity equations
should not depend on the site index and we can define the joint
probabilities $p_0=p(h=-2,\sigma=-1)$, $p_1 = p(h=-2,\sigma=1)$, $p_2
= p(h=0,\sigma=-1)$, ... , $p_5 = p(h=2,\sigma=1)$ of
Eq.~({\ref{iteration}) identically for all the lattice sites. A simple
enumeration shows that Eq.~(\ref{iteration}) reduces to the following
system of equations:
\begin{eqnarray}
\label{eq_system}
cp_0 e^{2\beta} = p_1^2 & 
cp_1 e^{-2\beta}= p_1^2 +2p_1p_3 + p_3^2 &
cp_2 = 2p_1p_2+2p_1p_4\nonumber\\
cp_3 = 2p_1p_4+2p_3p_4 &
cp_4  e^{-2\beta} = p_2^2 +2p_1p_3 + p_4^2&
cp_5 e^{2\beta} = p_4^2,
\end{eqnarray}
together with the normalization condition $\sum_{i=0}^5 p_i=1$. The
symmetries of the system fix some conditions on the values of
$p_i$. The relations $p_0 = p_5$, $p_1 = p_4$, $p_2 = p_3$ hold in the
paramagnetic phase (PA). If the overall $Z_2$ symmetry is spontaneously
broken the system encounters a ferromagnetic (FM) phase characterized by 
$p_2=p_3=p_4=p_5=0$ (obviously there is also the solution with the opposite magnetization, for which $p_0=p_1=p_2=p_3=0$). In the FM case the solution is given by 
\begin{equation}
\label{eq_fm_sol}
p_0 = \frac{1}{1+e^{4\beta}}\,\,\,\,\, p_1 = \frac{1}{1+e^{-4\beta}}
\,\,\,\,\, 	c = \frac{1}{e^{-2\beta}+e^{-6\beta}}.
\end{equation}
The solution in the PA case is slightly more involved:
\begin{eqnarray}
\label{eq_p_iter_hit}
&p_0& = \frac{c(x) D(x)^4 e^{-2\beta}}{16x^2}\,\,\,\,\,\,  
p_1 = \frac{c(x) D(x)^2}{4x}\,\,\,\,\,\,  
p_2 = \frac{c(x) D(x)^3}{2x}\nonumber\\
&c(x)& =
\frac{1}{2}\left[ \frac{D(x)^4}{64x^5}+\frac{D(x)^3}{8x^3}+\frac{D(x)^2}{4x}\right],
\end{eqnarray}  
where $x=(\frac{e^{2\beta}}{4})^{\frac13}$ and $D(x) = \sqrt{x^4 - 2x}
- x^2$. The problem of the iterative stability of both FM and PA
solutions can be addressed studying the spectrum of the Jacobian of the system of equations (\ref{eq_system}). It turns out that PA solution becomes unstable
below a $T_c^{PA}= \log^{-1}(2\sqrt{\frac23})=2.039091$ and the FM one
above $T_c^{{FM}}= \frac{2}{\log(2)}=2.885390$. Once we have
calculated the cavity fields we can measure the thermodynamic
potential using the merging procedures explained in the former
section. 
\begin{figure}
\begin{center}
\includegraphics[angle=0,width=0.9\textwidth]{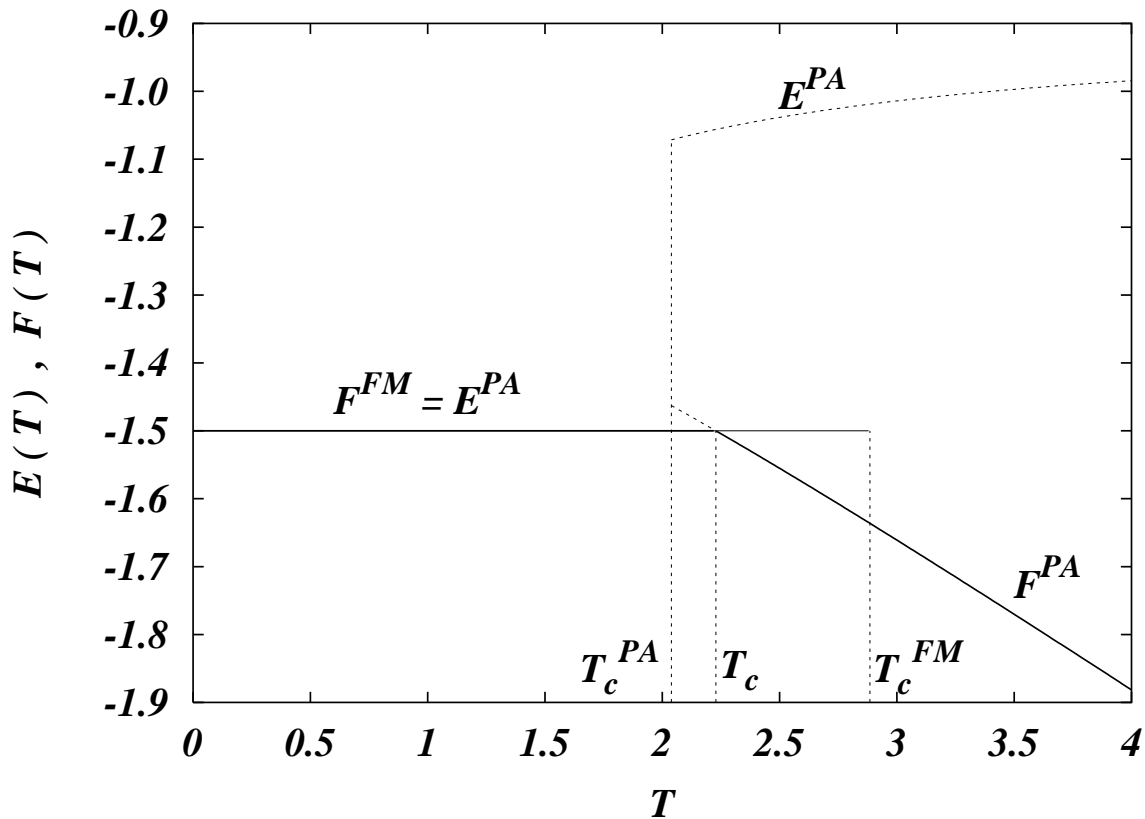}
\caption{Thermodynamic potentials $E$ and $F$ {\sl vs.} $T$ for the ferromagnet in the case $k=2$}
\label{fig_thermo}
\end{center}
\end{figure}

Let us consider first the FM solution. 
\begin{eqnarray}
\label{eq_ef_lowt}
\Delta E_1 = -3&& \Delta F_1 = -3\left[1-\frac1\beta\log(1+e^{-4\beta})\right]\nonumber\\
\Delta E_2 = -1&& \Delta F_2 = \frac2\beta\log(1+e^{-4\beta}) - 1.
\end{eqnarray}
Inserting Eq.~(\ref{eq_ef_lowt}) into Eq.~(\ref{F}) and
Eq.~(\ref{E}) we readily obtain $E = F = -\frac32$, {\em i.e.} the
system is completely frozen in its ground state.

The PA solution is worked analogously.
\begin{eqnarray}
\label{eq_ef_hit}
\Delta E_1 = -\frac{N_1(x)}{D_1(x)} && \Delta F_1 =
-\frac{1}{\beta}\log(D_1(x))\nonumber\\
\Delta E_2 = -\frac{N_2(x)}{D_2(x)} && \Delta F_2 =
-\frac{1}{\beta}\log(D_2(x)),
\end{eqnarray}
where 
\begin{eqnarray}
N_1(x) &=&  2p_1^3 (3e^{\beta}+e^{3\beta}) + 6 p_1^2p_2(e^{3\beta}+2
e^{\beta}) + 6p_1p_2^2(e^\beta+e^{3\beta})\nonumber\\
D_1(x) &=& 6 p_1^3(e^\beta+e^{3\beta})+  6 p_1^2 p_2(3e^{3\beta}+2e^{\beta}) 
+ 6p_1p_2^2(e^\beta+3e^{3\beta}) + 6p_2^3e^{3\beta}\nonumber\\
N_2(x) &=&  2 p_1^2 e^{-\beta} - 2 (p_1+p_2)^2 e^\beta\nonumber\\
D_2(x) &=&  2 p_1^2 e^{-\beta} + 2 (p_1+p_2)^2 e^\beta, 	
\end{eqnarray}
and for the $p_i$ we use Eq.~(\ref{eq_p_iter_hit}). In
Fig.~(\ref{fig_thermo}) we display both energy and free energy as a
function of the temperature. Starting in the low temperature phase, the system gets
trapped in the ferromagnetic solution in which all spins are
aligned. This solution is locally stable for the iteration
Eq.(\ref{eq_system}) up to temperature $T_c^{FM}$. We define
$T_c=2.228512$ as the temperature at which PA and FM solutions have
the same energy. Exactly at this temperature we have the coexistence
of the two phases of the system. Above $Tc$ the PA solution acquires a
lower free energy. In Fig.~(\ref{fig_entropy}) we display the
configurational entropy $S_C(E)$ as a function of
the energy $E$. The thin straight line is tangent to the curve exactly
at $S_C(E_c)$ where $E_c \equiv E(T_c)=-1.05613$, while the dotted curve
beneath the tangent is the result of the microcanonical computation of
Dean et al. presented in \cite{lefevredean}. Note that the
microcanonical branch can not be obtained in our canonical scheme.

\begin{figure}
\begin{center}
\includegraphics[angle=0,width=0.9\textwidth]{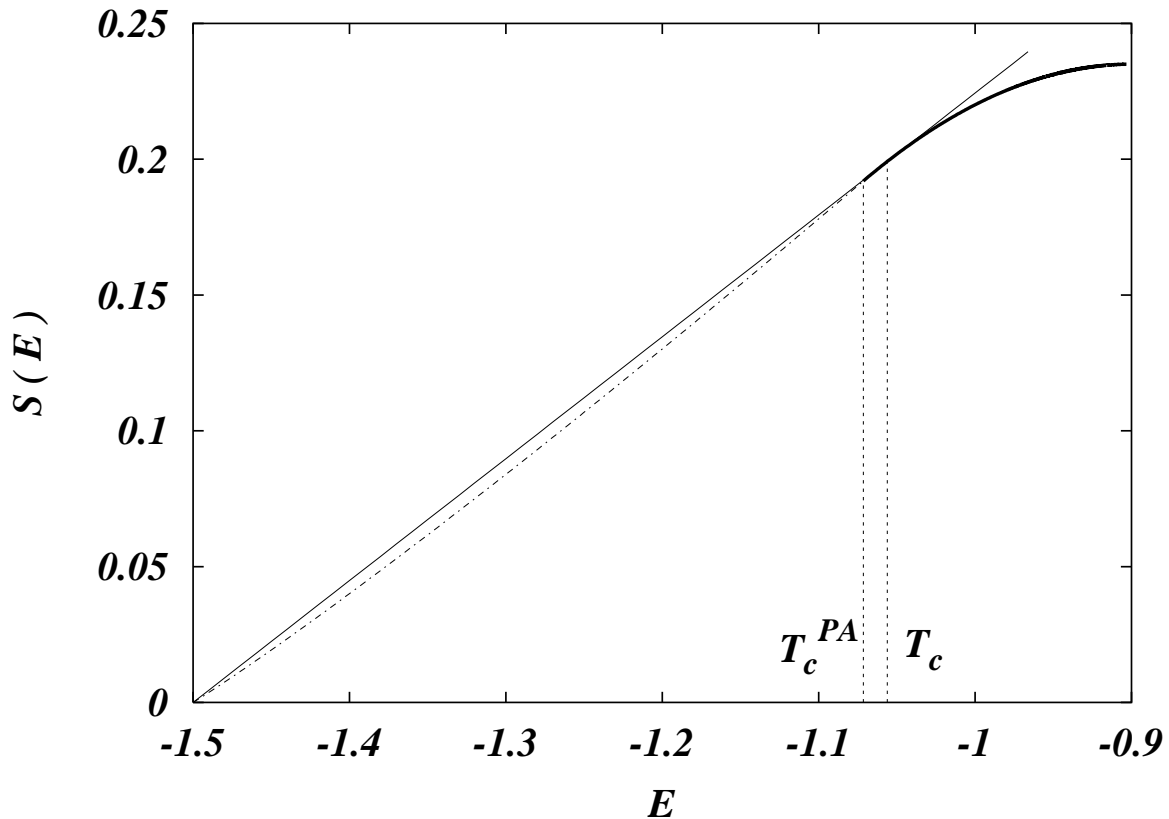}
\caption{Configurational entropy $S_C$ {\sl vs.} 
$E$ for the ferromagnet in the case $k=2$ (see text).}
\label{fig_entropy}
\end{center}
\end{figure}

\section{The binary spin glass}
\label{binary}

Let us now turn to a slightly more complicated case, the binary spin
glass, where the coupling constants are quenched independent
identically distributed (i.i.d.) random variables with the following
law:
\begin{equation}
\label{eq_pj}
\lambda(J)=\frac{1}{2}(\delta(J-1)+\delta(J+1)).
\end{equation}
As in the above section, an important simplification occurs: the local
field acting on a spin can have only a finite number of values. The
cavity fields can have $k+1$ values: $-k,-k+2,\dots,k$, whereas the
local field acting on a site with $k+1$ first neighbors can have
$(k+2)$ values: $-k-1,-k+1,\dots k+1$. But by contrast to the
ferromagnet, the system is no longer homogeneous, and the self
consistency equation (\ref{selfcon}) is a tricky object. To tackle a
similar equation, \cite{mezpar} suggested a numerical solution {\sl
via} a population algorithm. The basic idea is that a probability
distribution can be represented by a large collection of i.i.d. random
variables distributed according to it.

In our case, the algorithm works as follows: we use a large population
of $\cN$ sites $i=1,\dots \cN$, each of which is characterized by a
finite set of $2(k+1)$ numbers, the $p_i(h,\s)$. After random
initialization, one iterates the following sequence:
\bigskip

{\bf Algorithm I} 

\begin{itemize}
\item[(i)] Select $k+2$ sites $i_1,\dots i_{k+2}$ at random, and
extract some couplings $J_1,\dots J_{k+1}$.
\item[(ii)] Perform the merging (M1) of the branches rooted at
$i_1,\dots i_{k+1}$ onto the site $\Psi=i_{k+2}$, that is compute
$\Delta F_1$ with Eq.~(\ref{dF1}), $\Delta E_1$ with Eq.~(\ref{dE1}).
\item[(iii)] Perform the merging (M2) of the branches rooted at $i_1$
and $i_2$, that is compute $\Delta F_2$ with Eq.~(\ref{dF2}), $\Delta
E_2$ with Eq.~(\ref{dE2}), using for instance the coupling $J_1$.
\item[(iv)] Update the population: perform the merging of the branches
rooted at $i_1,\dots i_{k}$ onto the site $\Phi=i_{k+2}$, that is
substitute $p_{k+2}(h,\s)$ by the result of Eq.~(\ref{iteration}).
\end{itemize}

The algorithm converges in a stochastic sense: after a sufficient
number of iterations, the $p_i(h,\s)$ are distributed according to
$\cP$. So one can compute the average of the thermodynamic quantities
with respect to the couplings by 'time' averaging of the measures
performed in (ii) and (iii).

In the practical implementation of the algorithm we have scanned
values of $\cN$ ranging from $200$ to $4000$. A careful finite size
scaling analysis shows that results are really mildly dependent on
$\cN$ as soon as $\cN>1000$, and the asymptotic extrapolation is
always compatible, within statistical errors, with the biggest size we
have simulated. The errors are calculated with standard binning
procedure, discarding the first half of the simulation. In this way we
have a complete control on the ``{\em thermalization}'' of the
algorithm.  Most of the simulations have been performed also for the
same system removing the stability condition on the local fields. We
will always refer to these data as AC (all configurations) to
distinguish them from the metastable (MS) set of data. For most of the
simulations we have performed from $10^3 \times\cN $ up to $10^4
\times \cN$ iterations of Algorithm {\bf I}, which we have
verified to be enough both for {\em thermalization} and numerical
accuracy.

The configurational entropy $S_C$ is the MS entropy.

\begin{figure}
\begin{center}
\includegraphics[angle=0,width=0.9\textwidth]{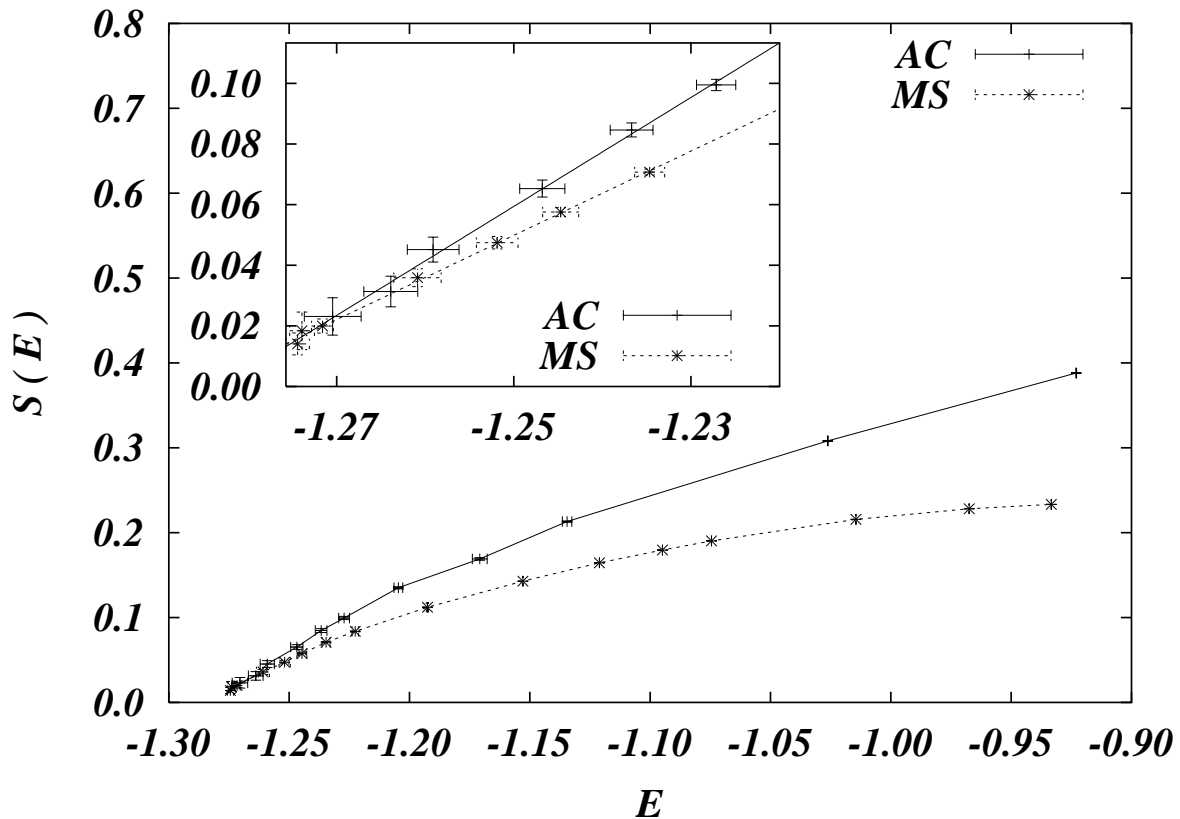}
\caption{ Entropy {\sl vs.} $E$ for the binary spin glass in the case
$k=2$. Inset: zoom of the low-energy region. MC with, AC without
one spin-flip stability condition.}
\label{fig_sde_bin}
\end{center}
\end{figure}

One should also note that in the case of the Bethe lattice the choice
of the distribution Eq.~(\ref{eq_pj}) is not crucial and the same
results hold in the case of a distribution
$\lambda(J)=p\,\delta(J-1)+(1-p)\delta(J+1)$ regardless how small $p$
is and even in the limiting case of a purely anti-ferromagnetic
system. This observation allows for the analytic computation of the
critical temperature and the critical energy of this model. Following
the route specified in Sec.~\ref{sec_fm} we calculated the equivalent
of the system of equations~(\ref{eq_system}) for purely
anti-ferromagnetic couplings. The stability analysis of the
paramagnetic solution gives the same temperature of the FM case, {\em
i.e.} $T_c = \log^{-1}(2\sqrt{\frac23})$ and $E_c \equiv E(T_c) =
-15/14 \approx -1.071429$. Below this temperature (energy) the replica
symmetry is spontaneously broken.

In figure (\ref{fig_sde_bin}) we display the logarithm of the number
of metastable states as a function of the energy. In the inset we zoom
the results for the lowest energy. It is interesting to note that in
the region $E<-1.22$ both AC and MS entropy seem to behave linearly on
$E$, and a linear fit works really well. The two curves meet at
$E=-1.273(5)$ and $S= 0.0172(5)$. This energy is clearly compatible
with the replica symmetric value of the ground state $E_0^{RS} =
-23/18 = -1.277777$ \cite{PARISI_MEZARD_86}. We have an excess of
entropy at $E_0^{RS}$ which is proportional to the density of {\em
spin fous} in the ground state, i.e. the number of spins with zero
local field -- such a spin can be flipped without changing the energy.

\begin{figure}
\begin{center}
\includegraphics[angle=0,width=0.9\textwidth]{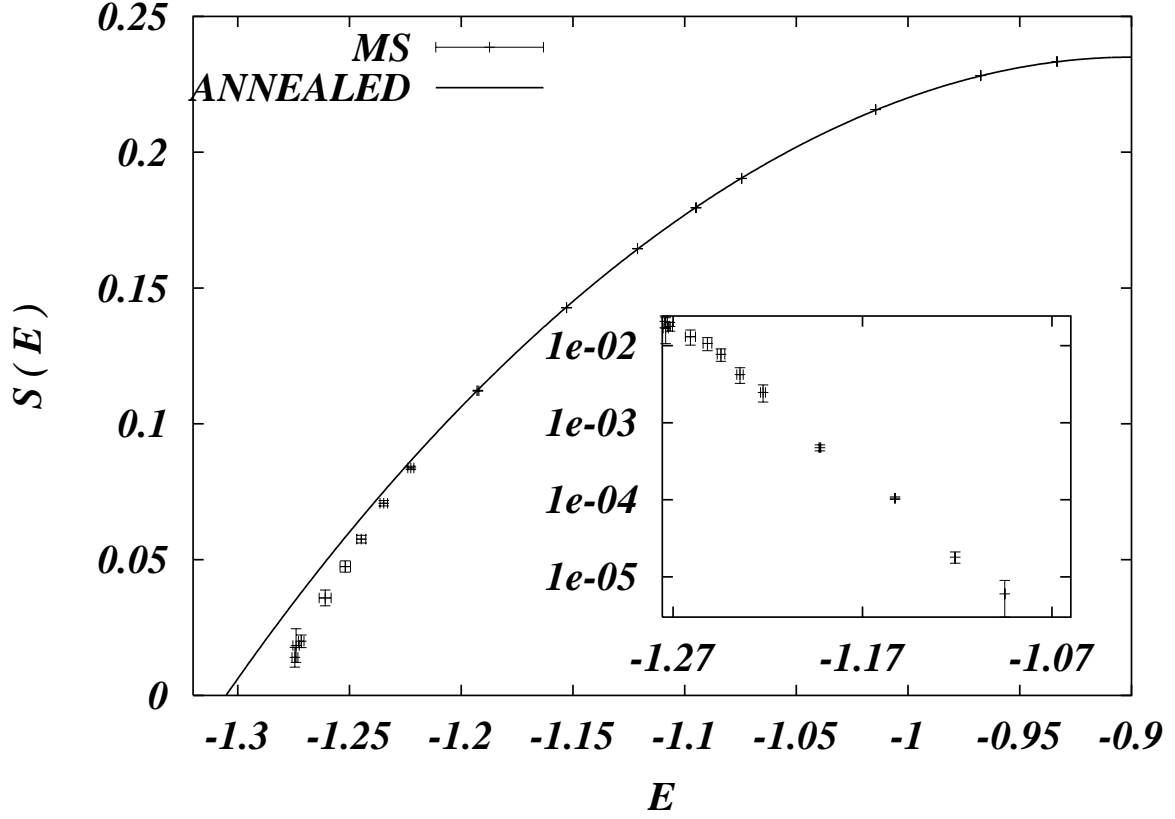}
\caption{ Entropy {\sl vs.} $E$ for the binary spin glass in the case
$k=2$: comparison of the quenched and annealed results. Inset: zoom of
the difference between the annealed and metastable (MS) entropy.}
\label{fig_anneal}
\end{center}
\end{figure}

In figure (\ref{fig_anneal}) we compare our data with the annealed
approximation presented in \cite{lefevredean}, while in inset we zoom
the difference $S_{ann}(E) - S(E)$ in the region $E<E_c$. Above this
value the annealed approximation is believed to be exact, and in fact
our data fall on the analytic curve within the error bars, while below
$E_c$ the two curves split. It is interesting to note that the
splitting, barely visible in the main panel for $E > -1.2$, seems to
be exponential, at least near $E_c$ (note the logarithmic scale on the
$y$ axis of inset).

Let us stress here that it is straightforward to adapt the algorithm
to the case of random graphs of fluctuating connectivity of finite
mean $c$. In such a graph, given two points, there is an edge
connecting them with probability $c/N$, and no edge with probability
$1-c/N$. Thus the number of first neighbors of a given site is a
random variable distributed according to a Poisson law of mean $c$. To
implement this, one must extract the number $k+1$ according to this
law in step (i). The rest is unchanged, except the important fact that
step (iv) is no longer to be performed with $k$ branches, but with
$k+1$ branches, like (ii) and that the $k+1$ term in Eqs.~(\ref{F})
and (\ref{E}) should be replaced by $c$ \cite{MEZ_ZEC}.

\section{The Gaussian spin glass}
\label{gaussian}

Our method can tackle even more complicated cases, such as the
Gaussian spin glass, in which the coupling constants are quenched
i.i.d random variables whose law is a Gaussian with unit variance and
zero mean.

The local fields are now continuous variables. We choose to represent
the probability distribution $p_i(h,\s)$ on a site by a large
population of $\cL$ couples $(h,\s)$ distributed according to $p_i$.

So on each site $i=1,\dots \cN$, we have a population
$(h_i^\nu,\s_i^\nu), \nu=1,\dots \cL$. The algorithm's layout is
similar to algorithm I:

\bigskip
{\bf Algorithm II}

\begin{itemize}
\item[(i)] Select $k+2$ sites $i_1,\dots i_{k+2}$ at random, and
extract some couplings $J_1,\dots J_{k+1}$.
\item[(ii)] Perform the merging of the branches rooted at $i_1,\dots
i_{k+1}$ onto the site $\Psi=i_{k+2}$. In the context of this
algorithm, the Eq.(\ref{dF1}) and (\ref{dE1}) become
\begin{equation}
\Delta F_1=-\frac{1}{\beta} \ln \[[\sum_{\nu=1}^\cL\sum_{\s=\pm 1} \,e^{\beta H_\Psi^\nu \s} \Theta(H_\Psi^\nu \s) \prod_{j=1}^{k+1} \Theta \[[(h_{i_j}^\nu+J_{j} \s)\s_{i_j}^\nu\]] \]]
\end{equation}
\begin{equation}
\Delta E_1=-\frac{\sum_{\nu=1}^\cL\sum_{\s=\pm 1} \,H_\Psi \s e^{\beta H_\Psi^\nu \s} \Theta(H_\Psi^\nu \s) \prod_{j=1}^{k+1} \Theta \[[(h_{i_j}^\nu+J_{j} \s)\s_{i_j}^\nu\]]}{\sum_{\nu=1}^\cL\sum_{\s=\pm 1} \, e^{\beta H_\Psi^\nu \s} \Theta(H_\Psi^\nu \s) \prod_{j=1}^{k+1} \Theta \[[(h_{i_j}^\nu+J_{j} \s)\s_{i_j}^\nu\]]},
\end{equation}
where $H_\Psi^\nu=\sum_{j=1}^{k+1} J_j \s_{i_j}$.
\item[(iii)] Perform the merging of the branches rooted at $i_1$ and
$i_2$, using for instance the coupling $J_1$. The formulas (\ref{dF2}) and
(\ref{dE2}) read: 

\begin{equation} 
\Delta F_2=-\frac{1}{\beta} \ln
\[[\sum_{\nu=1}^\cL \,e^{\beta J_1 \s_{i_1} \s_{i_2}} \Theta
\[[(h_{i_1}^\nu+J_1 \s_{i_2}^\nu)\s_{i_1}^\nu\]]\, \Theta \[[
(h_{i_2}^\nu+J_1 \s_{i_1}^\nu)\s_{i_2}^\nu\]] \]] \ee \be \Delta
E_2=-J_1\frac{\sum_{\nu=1}^\cL \,\s_{i_1} \s_{i_2}e^{\beta J_1
\s_{i_1} \s_{i_2}} \Theta \[[(h_{i_1}^\nu+J_1
\s_{i_2}^\nu)\s_{i_1}^\nu\]]\, \Theta \[[ (h_{i_2}^\nu+J_1
\s_{i_1}^\nu)\s_{i_2}^\nu\]]}{\sum_{\nu=1}^\cL \,e^{\beta J_1 \s_{i_1}
\s_{i_2}} \Theta \[[(h_{i_1}^\nu+J_1 \s_{i_2}^\nu)\s_{i_1}^\nu\]]\,
\Theta \[[ (h_{i_2}^\nu+J_1 \s_{i_1}^\nu)\s_{i_2}^\nu\]]}.  
\end{equation}
\item[(iv)] Update the population, performing the merging of the
branches rooted at $i_1,\dots i_{k}$ onto the site $\Phi=i_{k+2}$: one
wants to substitute the $(h_\Phi^\nu,\s_\Phi^\nu),\nu=1,\dots \cL$
with a new population distributed according to $p_\Phi(h,\s)$ of
Eq.~(\ref{iteration}). This is performed in two steps
\begin{itemize}
\item First one builds a list
$(\tilde{h}^{\nu},\tilde{\s}^{\nu}),\nu=1,\dots \cL$ as
follows. Start with $\nu=1$. Enter a loop in which you extract
randomly an $\nu'\in\{1,\dots \cL\}$ and a spin value $\s=\pm 1$,
until you have the stability condition $\forall j\in\{1,\dots k\},
(h_{i_j}^{\nu'}+J_{j} \s)\s_{i_j}^{\nu'}\geq 0$. Then set
$\tilde{h}^{\nu}=\sum_{j=1,\dots k} J_{j} \s_{i_j}^{\nu'}$ and
$\tilde{\s}^{\nu}=\s$. Increment $\nu$, and enter the loop again.
\item The list $(\tilde{h}^{\nu},\tilde{\s}^{\nu})$ is not the one to
overwrite the $(h_\Phi^\nu,\s_\Phi^\nu)$, because one must enforce the
presence of the factor $e^{\beta h\s}$ in Eq.~(\ref{iteration}). Thus the
elements of the list are to be reweighted: a suitable new population
$(h_\Phi^\nu,\s_\Phi^\nu),\nu=1,\dots \cL$ is obtained by repeating
$\cL$ times the process of picking an element in the list
$(\tilde{h}^{\nu},\tilde{\s}^{\nu}),\nu=1,\dots \cL$ with a
probability proportional to $e^{\beta
\tilde{h}^{\nu}\tilde{\s}^{\nu}}$.
\end{itemize}
\end{itemize}

\begin{figure}
\begin{center}
\includegraphics[angle=0,width=0.9\textwidth]{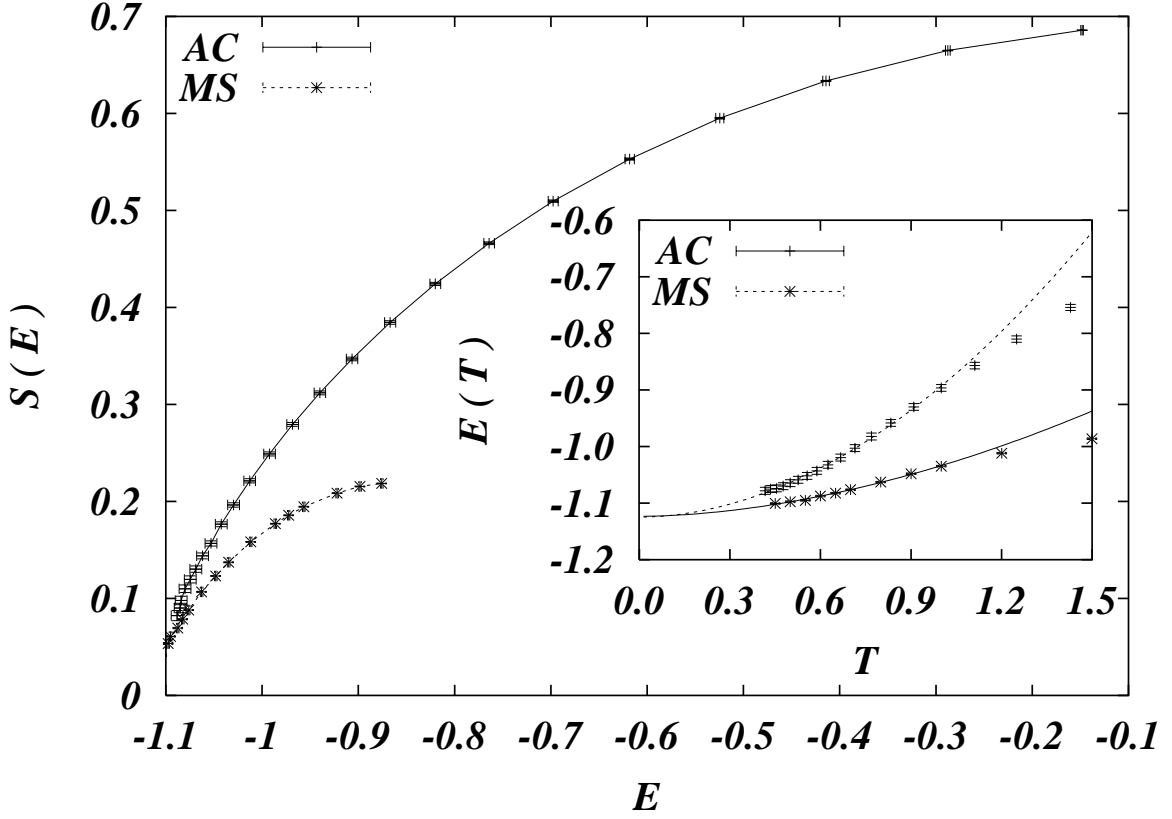}
\caption{ Entropy {\sl vs.} $E$ for the Gaussian spin glass. Inset:
Energy {\sl vs.} $T \equiv 1/\beta$. Lines are power-law fits of the
low-$T$ region. Note that both curves extrapolate at $T=0$ to the same
value (see text).}
\label{fig_sde_gau}
\end{center}
\end{figure}
We implemented the above algorithm in the case $k=2$, using the values
$\cN=1000$ and $\cL=1000$. The results are presented in
Fig.~(\ref{fig_sde_gau}). By contrast with the binary spin glass, the
ground state is not degenerated. So both entropies AC and MS go to 0
at the energy density of the ground state $E_{GS}$. Using the
algorithm of \cite{mezpar} at low temperature and extrapolating the
results to $T=0$, one finds $E_{GS}=-1.12(1)$, which is compatible
with our data.

\section{Breaking the replica symmetry}
\label{1RSB}

Let us now see how our method can be generalized to perform RSB.

\subsection{General considerations}

So far we have assumed there is a single pure state, stable under the
iteration process. It is well known that this is actually not true for
the spin glass on the Bethe lattice \cite{mezpar}. One should allow
for an infinity of pure states as described by the hierarchical
continuous RSB scheme proposed by Parisi \cite{mpv}. As a first
approximation, one can implement the 1RSB scheme. Briefly speaking,
one assumes that there exists an infinity of pure states labeled by
$\alpha=1,\dots +\infty$. The free energies of the states on one
branch are independent identically distributed random variables, with
an exponential density

\be
\rho(F)=\exp (\beta x (F-F^R))
\ee

where $F^R$ is a reference free energy, and $x$ is Parisi's parameter
\cite{mpv}. In this approach one computes the average values of the
observables as a function of $x$. The physical value of $x$ is the one
which maximizes the free energy.

If one considers a branch rooted at site $\Phi$, in each pure state
$\alpha$ the probability distribution $p_\Phi^\alpha(h,\s)$ is
different, and the iteration Eq.~(\ref{iteration}) is valid only
inside a given pure state. The self-consistency equation to be solved
is consequently more complicated than in the replica symmetric
case. Let us write ${\bf
p}_i(h,\s)=(p_i^\alpha(h,\s))_{\alpha=1,\dots}$. One is interested in
the set of probability distributions $\cQ_\Phi^{(h,\s)}({\bf
p})=\overline{\delta({\bf p}_\Phi(h,\s)-{\bf p})}$, which are
functionals of the analogous quantities $\cQ_i^{(h,\s)}$, and one asks
for all of them to be equal. A comprehensive theoretical description
of the extension of the cavity method to 1RSB can be found in
\cite{mezpar}.

Let us turn to the description of the algorithm. We use a population
of $\cN$ sites $i=1,\dots \cN$, $\cM$ states $\alpha=1,\dots
\cM$. On each site, each state is characterized by its own
distribution $p_i^\alpha(h,\s)$. Depending on the distribution of the
coupling constants, $p_i^\alpha(h,\s)$ is represented as a set of
$2(k+1)$ numbers (case of binary couplings), or again as a population of $\cR$
couples $(h,\s)$ (case of Gaussian couplings).

The RS algorithm {\bf I} is nested into the present algorithm: it is
used to implement the iteration (\ref{iteration}) and compute the
expectation values of the observables inside each state. Then a
meta-algorithm takes into account all the states, with appropriate
weights, to update the population and compute the global expectation
values. Two new observables are required with respect to the RS
case. First the variation of the free energy during the iteration process
described by Eq.~(\ref{iteration}):

\begin{equation}
\Delta F_{iter}=-\frac{1}{\beta} \ln \[[\mathop{\Tr}_{\s,\s_1,\ldots \s_k} \,e^{\beta h_\Phi\s} \prod_{i=1}^{k} \int d h_i \, \Theta \[[(h_i+J_{\Phi, i} \s)\s_i\]] p_i(h_i,\s_i) \]].
\end{equation}
Second the derivative of the free energy $F$ with respect to $x$. By
derivation of Eq.~(\ref{dF1}) and (\ref{dF2}), one gets \be
\frac{dF}{dx}=-\frac{F}{x}+d^1-\frac{k+1}{2} d^2, \ee where \be
d^1=\frac{1}{x} \frac{\sum_\alpha \Delta F_1^\alpha e^{-\beta x \Delta
F_1^\alpha}}{\sum_\alpha e^{-\beta x \Delta F_1^\alpha}}, \ee and
similarly for $d^2$.

An iteration of the algorithm goes as follows:
\bigskip

{\bf Algorithm III}
\begin{itemize}
\item[(i)] Perform step (i) of algorithm {\bf I}.
\item[(ii)] For each state $\alpha$, perform steps (ii), (iii) and
(iv) (computing {\sl en passant} $\Delta F_{iter}^\alpha$), of
algorithm {\bf I}. One gets quantities bearing $\alpha$ as a
superscript: $\Delta F_1^\alpha,\, \dots$.
\item[(iii)] Reweight the states: this is a crucial step, motivated in
\cite{mezpar}. The states with low $\Delta F_{iter}^\alpha$ have to be
favored. Thus one picks up $\cM$ times an element in the list of
distributions $p_\Phi^\alpha,\alpha=1,\dots \cM$ with a probability
proportional to $\exp(-\beta x\Delta F_{iter}^\alpha)$. The resulting
list overwrites the one at site $\Phi$. (Note that the elements one
picks up are composite objects, i.e. either a set of $2(k+1)$ numbers
or a population of $\cR$ couples).
\item[(iv)] Compute the global average values of the observables
$\Delta F_1 \dots$. The formulas can be found in \cite{mezpar}:
\begin{equation}
\Delta F_1= -\frac{1}{\beta x} \ln \[[ \frac{1}{\cM} \sum_\alpha
e^{-\beta x \Delta F_1^\alpha} \]]
\end{equation}

\begin{equation}
\Delta E_1= \frac{\sum_\alpha \Delta E_1^\alpha e^{-\beta x \Delta F_1^\alpha}}{\sum_\alpha e^{-\beta x \Delta F_1^\alpha}},
\end{equation}
and their obvious analogues for $\Delta F_2$ and $\Delta E_2$.

\end{itemize}
Note that it is essential that the sites and couplings be the same for
all the $\alpha$ in step (ii) above.

In practice, we found that the implementation of this algorithm in
the case of the Gaussian couplings requires too many computer
resources to reach a satisfying accuracy. So we limited ourselves to
the case of binary couplings.

\subsection{Application to the binary spin glass}

We considered the binary spin glass of section \ref{binary}, at
temperature $T=0.5$.  Our implementation used the values $\cN=1000$
and $\cM=1000$. We computed $dF/dx$ for several values of $x$. The
result is presented is Fig.~(\ref{fig_dfdx}). To determine the value
$x^*$ of $x$ where $dF/dx$ is zero, we fitted the curve by a
polynomial of degree 4, and found its roots. We got
$x^*=0.19(1)$. Then we measured the thermodynamic quantities for
$x=x^*$, taking into account the uncertainty on $x^*$:

\begin{equation}
F=-1.280 \pm 0.001,\,\,\,\,E=-1.265 \pm 0.001,\,\,\,\,S=0.029 \pm
0.001.
\end{equation}

This is to be compared to the output of the replica symmetric
algorithm of section \ref{binary}:

\begin{equation}
F=-1.2816 \pm 0.0008,\,\,\,\,E=-1.2685 \pm 0.0006,\,\,\,\,S=0.0274 \pm
0.0008.
\end{equation}

The overall improvement is small, particularly on $F$. It is more
obvious on $E$ and $S$ but still only of the order of $10^{-3}$. 

\begin{figure}
\begin{center}
\label{plotdF}
\includegraphics[angle=0,width=0.9\textwidth]{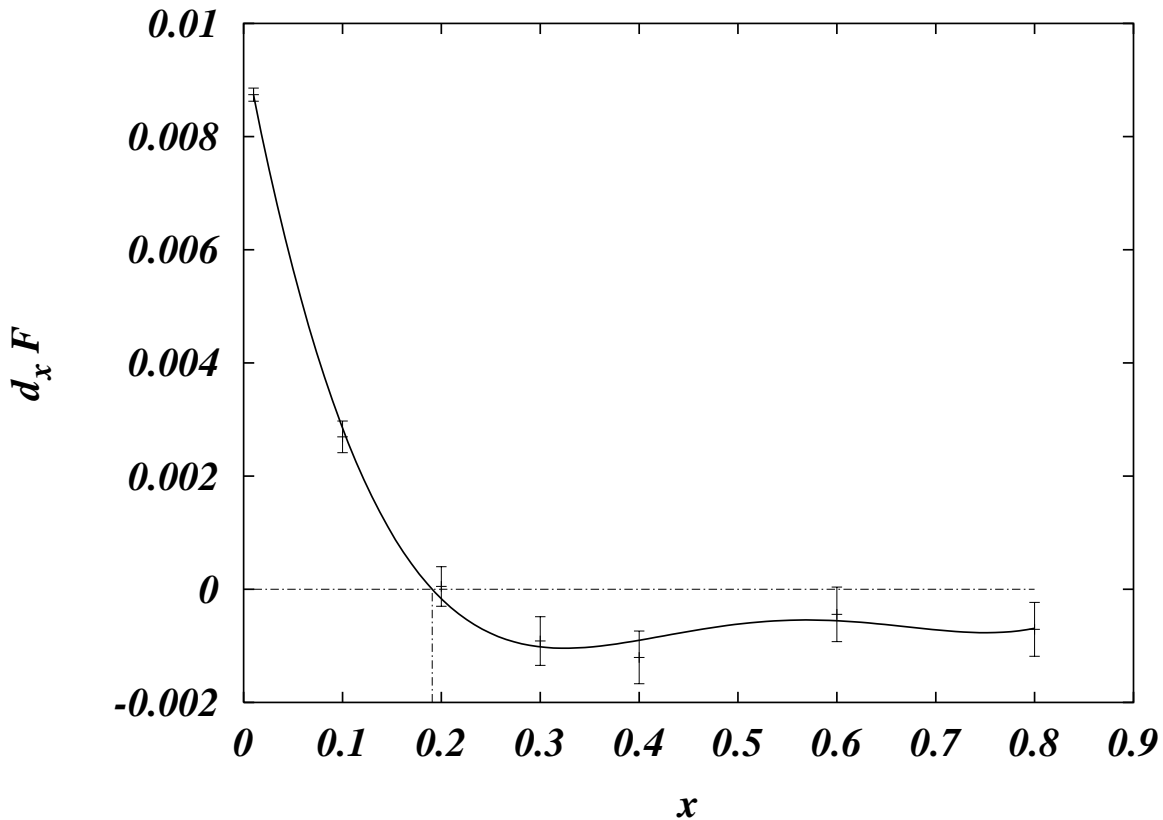}
\caption{$dF/dx$ {\sl vs.} $x$ for the binary spin glass in the case
$k=2$ at T=0.5; the continuous line is the polynomial fit of degree 4}
\label{fig_dfdx}
\end{center}
\end{figure}

\section{Conclusion}
In this paper we have described a general procedure to compute the
number of 1-spin flip stable configurations on the Bethe lattice.  Given
some integer $k \geq 2$, the method -- at least conceptually -- can be
easily generalized to compute the number of $k$-spin flips stable
configurations, i.e. whose energies can not be decreased by flipping a
number of spins ranging from 1 to $k$. The practical difficulty is
that the recursion relations can no longer involve only quantities
related to the root of a branch, but must also take into account the
$k-1$ successive generations of spins. In the case $k=2$ it remains
quite straightforward to write down the equations, but we have not
worked out their solution. It might be interesting to do so in order
to clarify the nature of the zero temperature limit of the pure
states: recently M\'ezard and Parisi \cite{PARISI_MEZARD_T0} presented
a computation at 1RSB level of the number of locally ground states
(LGS), i.e. configurations stable with respect to $k$-spin flips with
the number $k$ going to infinity with the size of the system in some unprecised way, and
they found some surprising features. There is a need for more precise
definitions, and one of the points at issue is how these LGS are
related to $k$-spin flips stable configurations when $k \rightarrow
+\infty$.
\begin{center}
{\bf ACKNOWLEDGMENTS}
\end{center}
We acknowledge very useful discussions with F. Ricci-Tersenghi.

\end{document}